\documentclass[%
 reprint,
 superscriptaddress,
 amsmath,amssymb,
 aps,
]{revtex4-2}

\usepackage{graphicx}
\usepackage{dcolumn}
\usepackage{bm}
\usepackage{hyperref}
\usepackage{xcolor}
\usepackage{soul}
\usepackage[normalem]{ulem}  

\begin{document}
\newcommand{\needed}{[{\color{magenta}citation}]}

\preprint{APS/123-QED}

\title{How do edge states position themselves in a quantum Hall graphene pn junction?}

\author{I.M. Fl\'or}
\affiliation{Univ. Grenoble Alpes, CEA, Grenoble INP, IRIG, Pheliqs, F-38000 Grenoble, France}
\author{A. Lacerda-Santos}
\affiliation{Univ. Grenoble Alpes, CEA, Grenoble INP, IRIG, Pheliqs, F-38000 Grenoble, France}
\author{G. Fleury}
\affiliation{Universit\'e Paris-Saclay, CEA, CNRS, SPEC, 91191, Gif-sur-Yvette, France}
\author{P. Roulleau}
\affiliation{Universit\'e Paris-Saclay, CEA, CNRS, SPEC, 91191, Gif-sur-Yvette, France}
\author{X. Waintal}%
 \email{xavier.waintal@cea.fr}
\affiliation{Univ. Grenoble Alpes, CEA, Grenoble INP, IRIG, Pheliqs, F-38000 Grenoble, France}%


\begin{abstract}
Recent experiments have shown that electronic Mach-Zehnder interferometers
of unprecedented fidelities could be built using a graphene pn junction in the quantum Hall regime. In these junctions, two different edge states corresponding to two different valley configurations are spatially separated and form the two arms of the interferometer. The observed separation, of several tens of nanometers, has been found to be abnormally high and thus associated to unrealistic values of the exchange interaction. In this work, we show that, although the separation is due to exchange interaction, its actual value is entirely governed by the sample geometry and independent of the value of the exchange splitting. Our analysis follows the lines of the classical work of Chklovski-Shklovskii-Glazman on electrostatically induced edge state reconstruction and includes quantitative numerical calculations in the experimental geometries. 
\end{abstract}

\maketitle

Electronic interferometers have been envisioned as the building blocks for quantum technologies with propagating states, i.e. flying quantum bits \cite{bauerle_coherent_2018}. Most of the early experiments have used GaAs/GaAlAs heterostructures which have exceptionally high mobilities  \cite{ji_electronic_2003,roulleau_2008}. Some geometries involve no magnetic field \cite{yamamoto_electrical_2012}, but one often works at high field in the quantum Hall effect (QHE) regime \cite{ji_electronic_2003,roulleau_2008}. There, the different interfering paths correspond to the  one-dimensional edge states of the QHE. Recently, using specific heterostructures designed to screen the electron-electron interaction, interference of anyons in the fractional QHE regime have been observed \cite{nakamura_direct_2020}. 

\begin{figure}[h!]
    \centering
    \includegraphics[width=\linewidth]{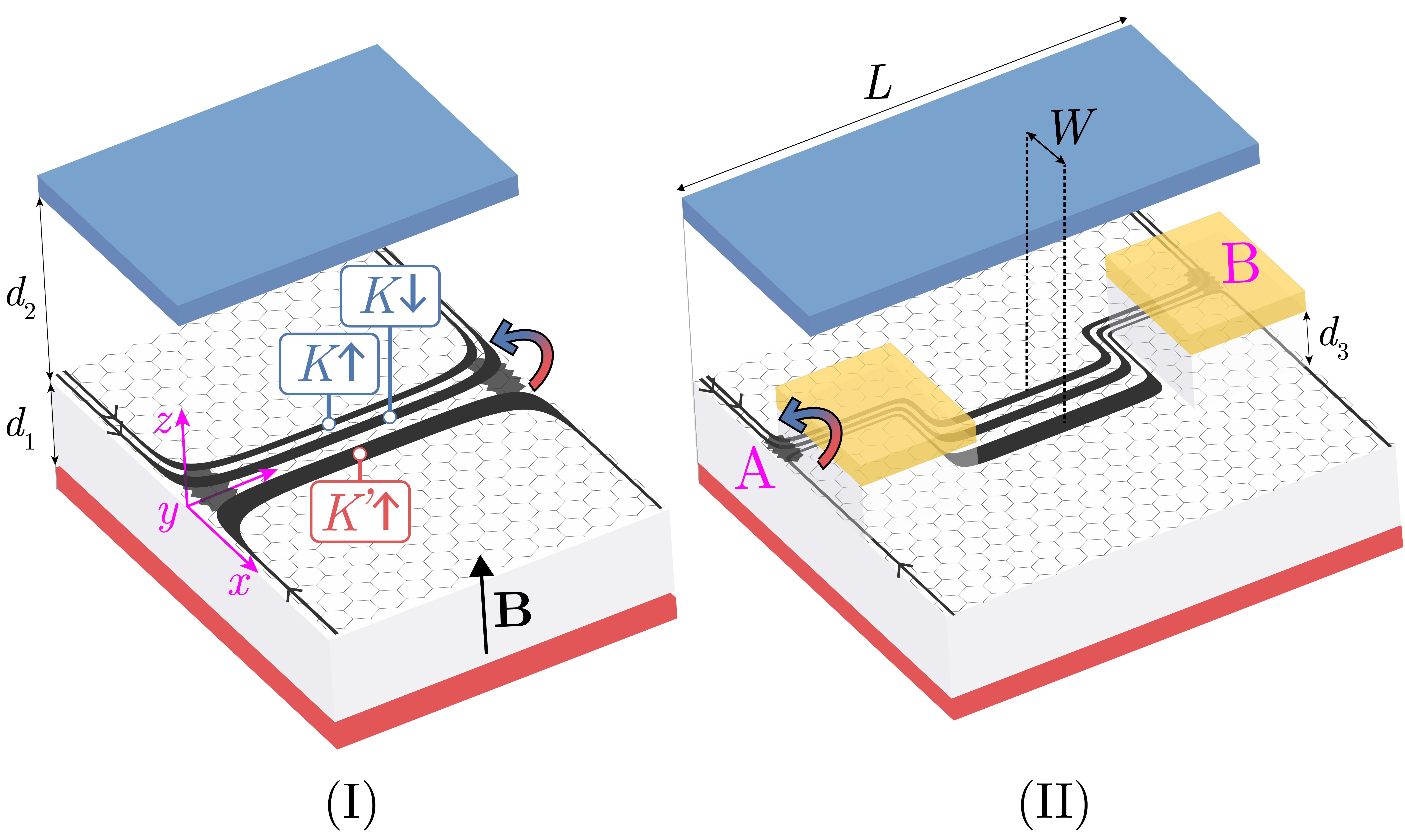}
    \caption{Two graphene pn junctions (I) and (II) respectively without \cite{wei_mach-zehnder_2017} and with \cite{jo_quantum_2021} side gates (in yellow). The top (blue) and bottom (red) gates create the pn interface. Insulating hBN fill the empty space between gates and graphene. A magnetic field $B=9$ T is applied in the $\hat{z}$ direction. Two spin-split quantum Hall channels with valley isospin ${K}$ propagate at the n side ($x<0$).  One spin-polarized channel with opposite valley isospin ${K}'$ propogates at the p side ($x>0$).}
    \label{fig:fig1}
\end{figure}

Graphene is another highly promising platform for these "electron quantum optics" experiments. The quality of graphene samples has improved drastically with two key techniques. First, the encapsulation with hexagonal boron nitride (hBN) \cite{Dean2010}. Second,  the usage of nearby graphite electrostatic gates that screen the electrostatic interaction as well as the charges trapped in the substrate at the Si/SiO2 interface. Graphene pn junctions based experiments now report record performances with interference visibilities nearing 100\% \cite{wei_mach-zehnder_2017, jo_quantum_2021}.

Most features of these graphene Mach-Zehnder experiments could be understood within a Landauer-B\"uttiker (LB) picture that we recall below. One observation, however, remained puzzling: the large separation $W$ between the two interfering channels that form the two arms of the Mach-Zehnder, ranging between $W=50$ and $200$ nm, depending on the experimental setup \cite{wei_mach-zehnder_2017, jo_quantum_2021}. 
In a naive LB picture, such a large separation should be associated with an abnormally high value of the exchange interaction, almost in the 100 meV range.
The purpose of this article is to show that this paradox is due to a breakdown of the LB picture in the QHE regime. 
The LB approach does not account for the (dominating) electrostatic energy \cite{armagnat_reconciling_2020} and must be replaced by the more elaborate Chklovski-Shklovskii-Glazman (CSG)  model \cite{chklovskii_electrostatics_1992}. 
 Performing the CSG construction of the compressible and incompressible stripes in a  graphene pn junctions, our main result is a simple explanation to the above mentioned paradox. 
 Even though the splitting between the interface states is indeed due to the presence of an exchange interaction, we find, in contrast to previous claims \cite{wei_mach-zehnder_2017}, that the actual value of the distance between the edge states is entirely controlled by the geometry  of the device (via its electrostatic properties) and is essentially independent of the value of the exchange splitting.

\paragraph*{Devices geometry and summary of the main experimental observations.} We consider the two geometries (I) and (II) displayed in Fig.~\ref{fig:fig1}.  They closely mimic the experimental setups used in 
Ref.~\cite{wei_mach-zehnder_2017} and Ref. \cite{jo_quantum_2021} respectively. We also consider a third geometry (III) studied in the supplementary material of \cite{jo_quantum_2021}. It is essentially identical to (I) but with a different value of the distance between the graphene layer and the top gate, noted $d_2$.
A hBN encapsulated graphene monolayer is sandwiched by two gates. 
A bottom gate (at voltage $V_b$) spans the full graphene flake while a top gate (at voltage $V_t$) is only present on half of the flake, $x<0$. 
By setting different values of the voltages $V_b$ and $V_t$, one may form a pn junction with e.g. electrons accumulated under the top gate and holes in the other part of the sample. 
A magnetic field $\mathbf{B}=+B\hat{z}$ is applied perpendicular to the graphene flake to bring the graphene layer into the QHE regime. Here, we focus on the situation with 
$V_b<0$ and $V_t>0$ such that the filling factor is $\nu=2$ at the n side and $\nu=-1$ at the p side, where $\nu = n_s h/(eB)$ ($n_s$: electron surface density).  In the n region, two channels circulate counter-clockwise while in the p region one propagates in the clockwise direction.
In setup (II), the two additional side gates (at voltage $V_s$) allow one to tune the transmission between the edge channels along the graphene boundaries and the interface states along the pn junction. 

For $-2\le \nu\le 2$, only a single Landau level is filled. This peculiar Landau level is pinned at the Dirac point (our energy reference) and is a specificity of the Dirac dispersion relation of graphene. In a non-interacting theory, this Landau level is degenerate in both spin ($\uparrow,\downarrow$) and valley ($K,K'$). The exchange interaction $E_{\rm ex}$, however, lifts this degeneracy \cite{werner_2020}. 
The existence of interference in these experiments relies on the intervalley scattering at the intersection between the physical edges of the sample and the pn interface \cite{tworzydlo_valley-isospin_2007}. 
In the LB picture, an incoming state --- say $K\uparrow$ coming from the n side at $y=0$ --- is scattered into a state $K'\uparrow$ (respectively $K\uparrow$) at point (A) in the pn junction with amplitude $S_{K'K}^A$ (respectively $S_{KK}^A$). 
The state then propagates along the pn junction (near $x=0$ between $y=0$ and $y=L$) as a superposition of $K\uparrow$ and $K'\uparrow$. It does so coherently due to valley conservation along the interface \cite{tworzydlo_valley-isospin_2007, trifunovic_valley_2019}. Then,  it is scattered again at the (B) corner with an amplitude $S_{KK'}^B$ (respectively $S_{KK}^B$) into an outgoing channel, say $K\uparrow$ towards the n side  where it further propagates towards an Ohmic contact situated at $x=-\infty$, $y=L$. Note that the valley index $K$, $K'$ is not necessarily well defined on the edges of the sample where intervalley scattering can occur (depending on the microscopic structure, say armchair versus zigzag) but we keep the same letter for labeling these states for convenience. The resulting differential conductance obtained from the Landauer formula is
\begin{equation}
\label{eq:LB}
g=\frac{e^2}{h}\left|S_{KK'}^B S_{K'K}^A+e^{i\Phi} S_{KK}^B S_{KK}^A\right|^2
\end{equation}
where the phase difference $\Phi = eBWL/\hbar$ is given by the magnetic flux accumulated along the interface between the two states $K\uparrow$ and $K'\uparrow$. Note that in this picture, the  $K\downarrow$ state has a different spin from the other two channels and is simply a spectator. 
Indeed, in the absence of magnetic impurities or spin-orbit coupling, spin is conserved along the edge states. Equation (\ref{eq:LB}) predicts that the conductance oscillates with magnetic field.  
The period of these oscillations directly provides the separation $W$ between the edge states $K\uparrow$ and $K'\uparrow$ in the pn junction, the length $L$ of the junction being defined by the sample geometry. 
The oscillations predicted by Eq.~(\ref{eq:LB}) are  observed experimentally with periods corresponding to large values of $W$.

\begin{figure}[h!]
    \centering
    \includegraphics[width=\linewidth]{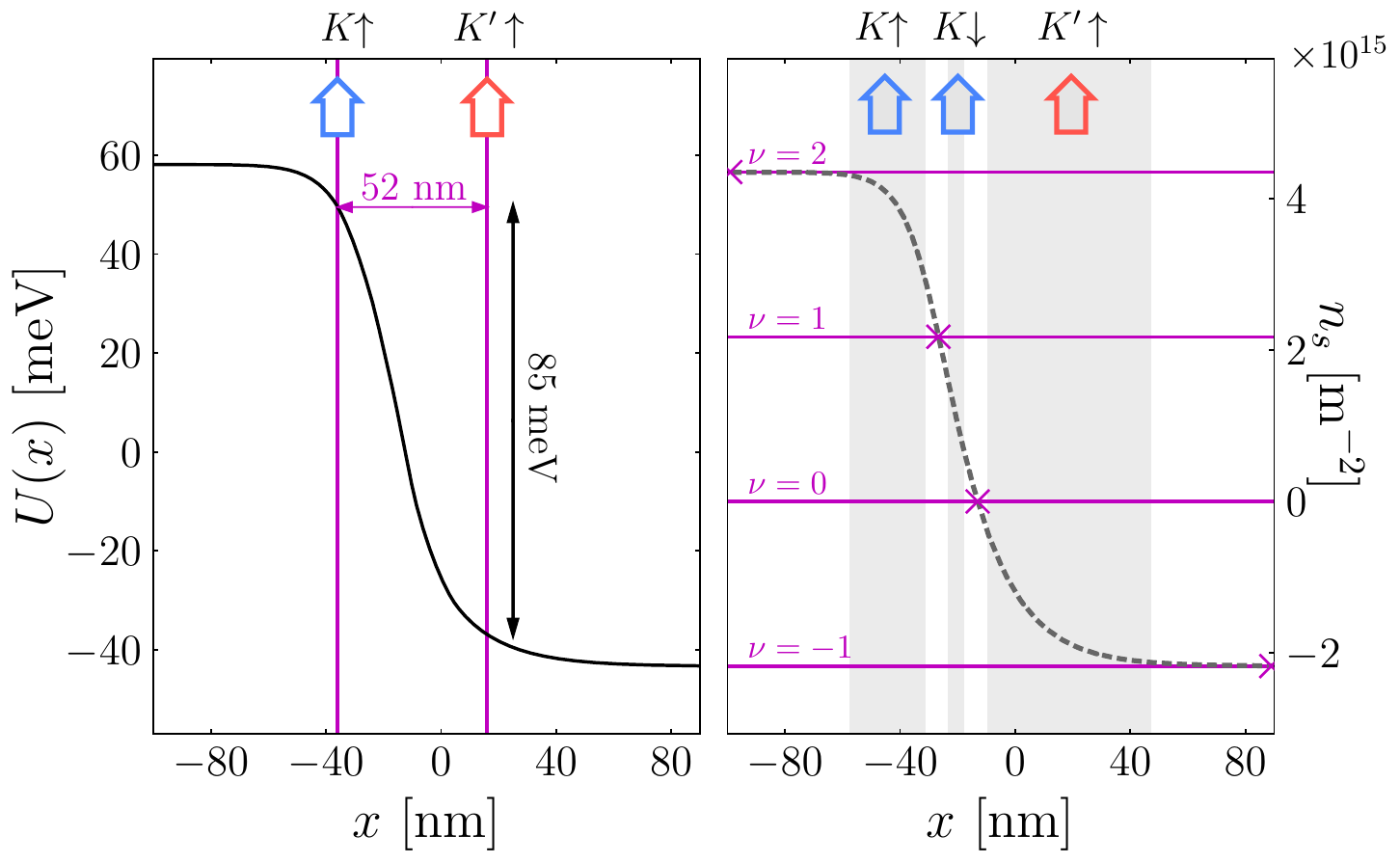}
    \caption{Calculations of a pn junction in the absence of magnetic field for the geometry (I) with $d_1=d_2=20$ nm
    as in Ref.~\onlinecite{wei_mach-zehnder_2017} and $V_b=-0.2$ V, $V_t=0.6$ V. (Left) Thomas-Fermi potential $U(x)$. Also shown is the value of $E_{\rm ex}\approx 85$ meV needed to account for the experimentally observed $W=52$ nm in \cite{wei_mach-zehnder_2017} according to the LB picture. (Right) electronic density $n_s(x)$ in the pure electrostatic approximation. Also shown are the positions in $x$ where the filling $\nu$ would take integer values at $B=9$ T. These locations will become incompressible stripes (in white) separating the conducting compressible stripes (in gray) following the CSG picture.  }
    \label{fig:fig2}
\end{figure}

\paragraph*{Qualitative role of the exchange interaction in the value of the edge state separation $W$.} 
We now focus on the pn junction and ignore the boundaries at $y=0$ and $y=L$. 
In the Landau gauge, a Landau level with momentum $k$ along the $y$-direction is centered along $x=k \ell_B^2$ where $\ell_B=\sqrt{\hbar/eB}$ is the magnetic length.  
In the presence of an electrostatic potential $U(x)$ that varies smoothly on the scale of $\ell_B$, the dispersion relation of the propagating channels (within the Dirac point Landau level) takes the form $E_p(k) = -eU(k \ell_B^2) + p E_{\rm ex}/4$. 
The integer $p\in\{\pm1,\pm3\}$ labels the $4$ different channels with different valleys $K,K'$ and spins $\uparrow,\downarrow$.
Here, the exchange interaction energy  takes the value  $E_{ex}$ for valleys and $E_{ex}/2$ for  spins. As we shall see, this choice of values will be mostly irrelevant in what follows. 
It follows from the dispersion relation that, for two channels at the Fermi energy $E_F$, the exchange energy must exactly compensate the change of electrostatic energy due to the spatial separation. Hence, 
for a constant gradient of potential,
\begin{equation}
e\frac{\partial U}{\partial x} W = E_{\rm ex}.
\label{eq:exc_halperin}
\end{equation}
In Equation (S1.6) of Ref.~\cite{wei_mach-zehnder_2017}, this equation was used to determine the width $W\approx E_{\rm ex}/ (e\partial U/\partial x)$. Below we argue that, while equation \eqref{eq:exc_halperin} is strictly speaking correct, it cannot be used to determine $W$. In contrast, it defines the value that the gradient of potential takes while $W$ is essentially determined by the geometry of the system (in the (I) geometry, $W \propto d_2$ the distance to the top gate).

\begin{figure}[t]
    \centering
    \includegraphics[width=\linewidth]{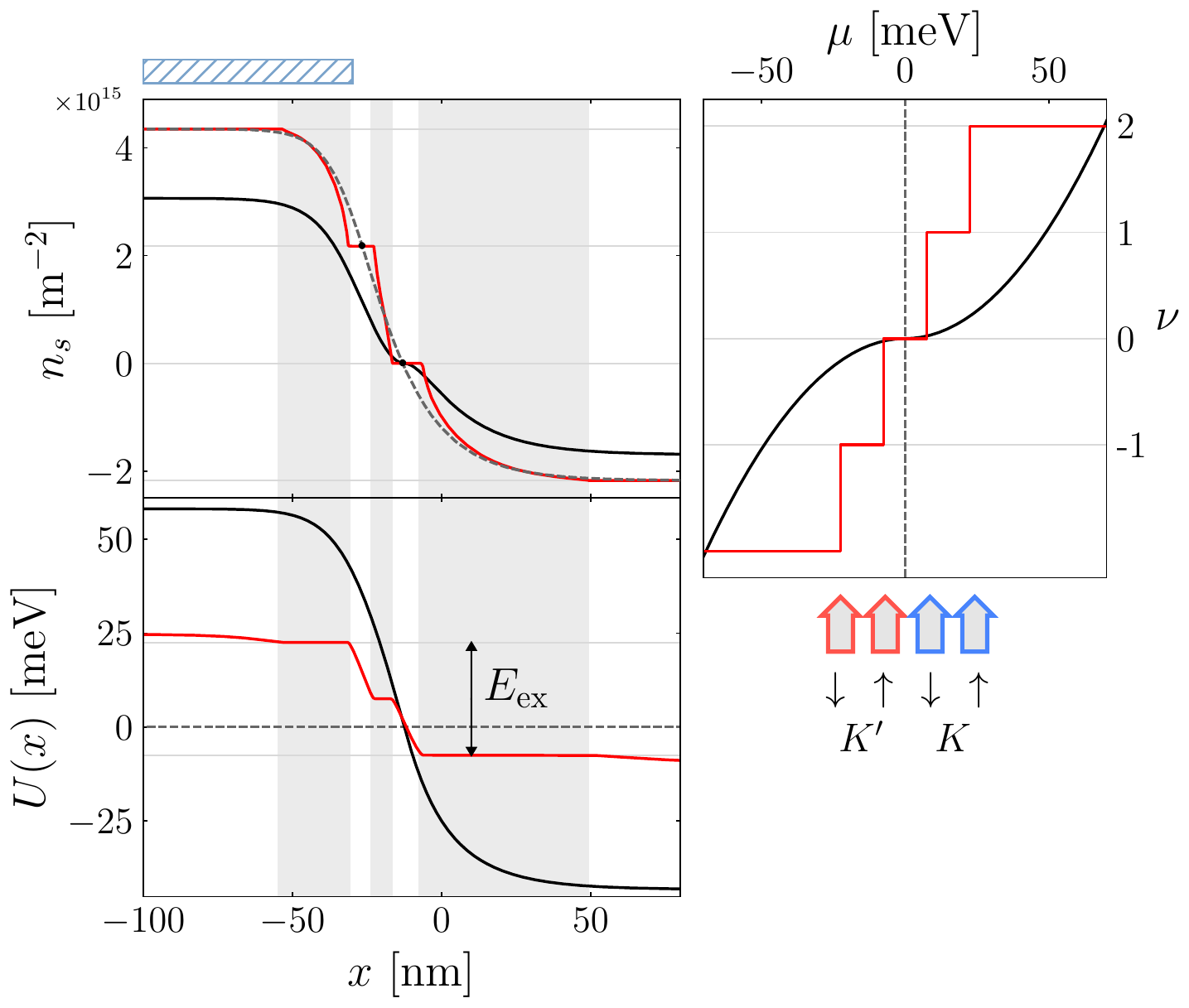}
    \caption{Reconstruction of the edge channels in geometry (I) with three approximations: pure electrostatic (PE) (dashed gray) and (generalized) Thomas-Fermi with $B=0$ (solid black) and $B=9$ T (solid red) respectively. For each case, the density profile (top left), the carrier density in the zeroth Landau level (top right) and the potential (bottom left) are shown. Here $d_1=d_2=20$ nm. The chosen value $E_\mathrm{ex}=30$ meV only affects the width of the incompressible region. The integer filling factors are shown with gray horizontal lines in both top plots. (In)compressible regions are shown in (white) gray patches and the horizontal position of the top gate is indicated by the blue hatched region.}
    \label{fig:fig3}
\end{figure}
\paragraph*{Construction of the Chklovski-Shklovskii-Glazman (CSG) compressible and incompressible stripes.} Following CSG \cite{chklovskii_electrostatics_1992}, we start by calculating the potential profile $U(x)$ and density profile $n_s(x)$ in the junction {\it in the absence} of magnetic field. 
We do so within the Thomas-Fermi (TF) approximation. In our 2D geometry (infinite pn junction along $y$), there is no bulk electronic density so that Poisson equation reads,
\begin{equation}
    \Delta U (\mathbf{r}) = 0
\end{equation}
with Dirichlet boundary conditions at the electrostatic gates. The graphene 2D electronic density $n_s(x)$ gives rise to a discontinuity of the electric field given by,
\begin{equation}
\frac{\partial U}{\partial z}(x,z=0^+) - \frac{\partial U}{\partial z}(x,z=0^-) = -\frac{e}{\epsilon} n_s(x)
\end{equation}
where $\varepsilon =4\varepsilon_0$ corresponds to the hBN dielectric constant. In the TF approximation, the density is controlled by the bulk graphene
density of state $\rho(E)$ which reads at zero temperature,
\begin{equation}
\label{eq:ildos}
n_s(x) = \int_0^{\mu=eU(x)} dE\ \rho(E),
\end{equation}
assuming that the Fermi energy $E_F=0=\mu-eU(x)$ ($\mu$: chemical potential) is constant across the graphene sheet. In the absence of magnetic field, graphene has a linear density of states
$\rho(E)=2|E|/\pi\hbar^2e^2v_F^2$ with $v_F\approx10^6$ m/s the Fermi velocity in graphene.
In the calculation shown in Fig.~\ref{fig:fig2}, we adjust the top and bottom gate voltages in order for the electronic density in the bulk n and p regions to correspond respectively to that of $\nu=2$ and $\nu=-1$ at $B=9$ T. Note, however, that the magnetic field remains zero in the calculation at this stage. The numerical calculation is performed using a generalization of the approach described in \cite{armagnat_self-consistent_2019-1}. 
In the left panel of Fig.~\ref{fig:fig2}, the electrostatic potential was calculated for the setup in experiment (I) where the distance between the top gate and the graphene layer is $d_2\approx20$ nm \cite{wei_mach-zehnder_2017}. 
Assuming that this potential profile would be weakly affected by the magnetic field, one finds [using Eq.~(\ref{eq:exc_halperin})] that the large value $W=52$ nm observed experimentally requires an exceedingly large exchange energy of $E_{\rm ex}\approx 85$ meV. 
As a reference, this value is almost as large as the distance to the next Landau level $\hbar v_F\sqrt{2}/\ell_B\approx 100$ meV. 
Such a large exchange energy would imply a deep reconstruction of the Landau levels that is not observed experimentally. 
The assumption that the electrostatic potential is unaffected by the magnetic field is in fact not valid \cite{armagnat_reconciling_2020}. 
In contrast, it is the electronic density $n_s(x)$ shown on the right panel of Fig.~\ref{fig:fig2} that is almost unaffected by the presence of a magnetic field. 
Indeed, modifying the electronic density can provide a gain in energy of the order of the exchange energy $E_{\rm ex}$ or the cyclotron frequency $\hbar\omega_c$ at a great loss in electrostatic energy.
This is favourable only when the density is close to an integer filling factor.  
Hence, in the spirit of the CSG approach, we identify the positions in the right panel of Fig.~\ref{fig:fig2} that correspond to integer values of $\nu$. Upon switching the magnetic field,
a small region around these points will become incompressible stripes with a flat density $n_s(x)=\nu h/eB$. Away from these points, $n_s(x)$ is not constant which means that there must be one partially filled Landau level pinned at the Fermi level. These regions are the compressible stripes, where propagation is allowed. In these regions, the electrostatic potential $U(x)$ remains constant. We refer to \cite{chklovskii_electrostatics_1992,chklovskii_ballistic_1993} for the details of the original construction and to \cite{armagnat_reconciling_2020,armagnat_self-consistent_2019-1} for a more recent version compatible with numerical calculations. In this picture, the size of each incompressible stripe is proportional to $\sqrt{E_\mathrm{ex}}$ \cite{armagnat_self-consistent_2019-1}. Their positions, however, are entirely determined by the electrostatic potential at $B=0$ hence by the geometry of the problem.
In particular the width $W$, that corresponds to the distance between the centers of the two outer compressible stripes, is entirely determined by the electrostatics (hence independent of $E_\mathrm{ex}$). Here we  estimate $W\approx 62$ nm, without adjustable parameter, which is in good agreement with the experimentally found value $W=52$ nm in experiment (I) for the same geometry.
For experiment (III) with $d_2=50$ nm, we find $W=90$ nm, also in good agreemeent with the value $W\approx 83$ nm found experimentally (see Fig.~S3 in the supplementary of Ref.~\onlinecite{jo_quantum_2021}). 

\paragraph*{Numerical calculations of the compressible/incompressible stripe structure.}
To actually calculate the stripes,  we now use the finite $B$ density of states. It is a sum of Dirac peaks at the positions of the Landau sublevels \cite{castro_neto_electronic_2009}.  

\begin{equation}
\rho(E) = \frac{1}{2\pi\ell_B^2} \sum_{n\in \mathbb{Z}} \sum_{\substack{p=\pm1\\,\pm3}} 
\delta\left(E -  E_n - \frac{p}{4} E_{\rm ex} \right)
\end{equation}
where $E_n=\hbar v_F\mathrm{sgn}(n)\sqrt{2|n|}/\ell_B$ are the Landau levels of degenerate graphene.
This Generalized Thomas-Fermi (GTF) approximation includes the effect of the (Fock) exchange interaction phenomenologically. Indeed, it is only the existence of a splitting and not its exact value that affects the results presented here.
At $B=E_{\rm ex}=0$, one recovers the TF approximation above. Considering only the $n=0$
Landau level in the limit $E_{\rm ex}=0$, we obtain the Pure Electrostatic (PE) approximation, i.e. the graphene is subject to a Dirichlet condition with an equipotential $U(x,z=0)=E_F=0$. 
The right panel of Fig.~\ref{fig:fig3} illustrates the three cases considered by showing the integrated density of state Eq.~(\ref{eq:ildos}) in  bulk graphene. The results of the self-consistent calculation are shown in the upper (density) and lower (potential) left panel of Fig.~\ref{fig:fig3}.
They are fully consistent with the picture described in the above paragraph. We have also verified (not shown) that the width of an incompressible strip is indeed proportional to $\sqrt{E_{\rm ex}}$ and that the value of $W$ does not depend on it. Hence the value of $E_{\rm ex}$ used in the calculations can be chosen arbitrarily. We note that the PE calculation approximates the quantum Hall graphene better than the $B=0$ TF one. This is unsurprising since the PE approximation naturally captures the position of the $n=0$ Landau level.

\begin{figure}[t]
    \centering
    \includegraphics[width=\linewidth]{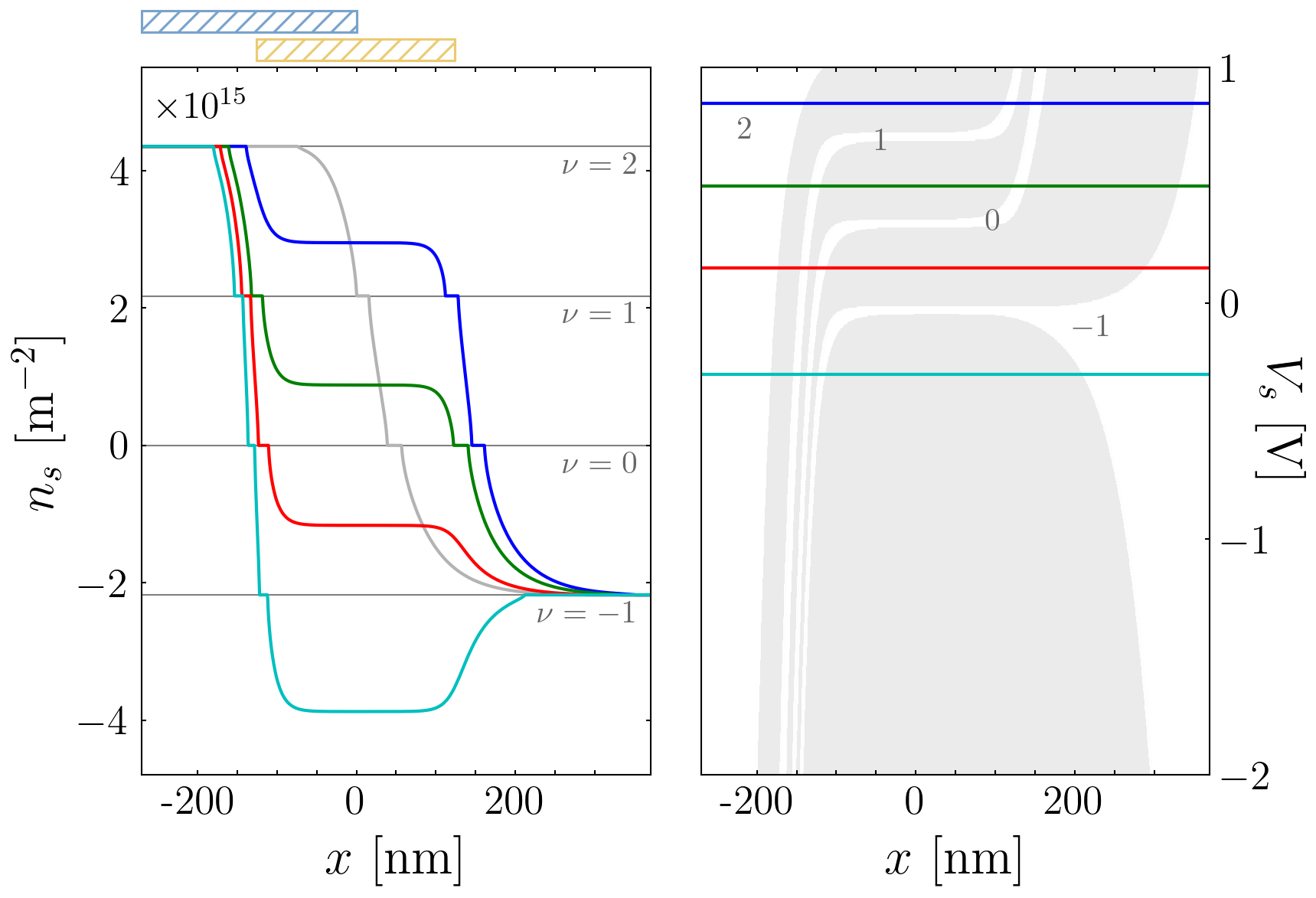}
    \caption{(Left) Density profiles calculated in the generalized Thomas-Fermi approximation for $B=9$ T and different values of $V_s$ (color) and for an identical device without side gate (gray). The top and side gate horizontal position are indicated by the hatched regions. (Right) (In)compressible stripes as a function of $V_s$ with lines drawn at the values for $V_s$ depicted in (left). $d_1=d_3=30$ nm and $d_2=60$ nm. Below $V_s=0$, the inter-channel separation decreases from 48 nm to 35 nm.}
    \label{fig:fig5}
\end{figure}
\paragraph*{Effect of a side gate.\label{sec:exp}}
We now turn to the experimental setup (II) with additional side gates at voltage $V_s$. The main usage of the side gates is to control the scattering amplitudes of equation (\ref{eq:LB}) in order to maximize the visibility of the interference pattern \cite{jo_quantum_2021}. We retain two experimental findings associated with this side gate: (i) The interference is only present for negative values of $V_s<-0.3$ V as shown in Ref.~\onlinecite{jo_quantum_2021} Fig.~2(a), 
(ii) the period of the oscillations is equal to 25 mT. It is roughly constant except close to $V_s=-0.3V$ where it is about 45\% smaller, around 14 mT
(values extracted from an analysis of the data of Fig.~3(b) in Ref.~\onlinecite{jo_quantum_2021}, the qualitative period change is visible with the bare eye). 

In Fig.~\ref{fig:fig5} (left), we distinguish four density profiles for different values of the potential $V_s$. When $V_s<0$, all compressible stripes are situated at the left ($x<125$ nm) part of the side gate. 
However, when $V_s>0$, an incompressible region necessarily finds itself extended over the entire width of the side gate in between two Landau sublevels. 
The inter-channel separation in this case is increased dramatically. Consistent with observation (i), inter-channel scattering at the graphene edges A and B is expected to be fully suppressed. 
As for observation (ii), $W$ steeply decreases as $V_s$ tends away from zero, resembling the experiment. Quantitatively, the experimental results correspond to  an average shift $dW/dV_s = 10$  nm/V; in our calculations, the center of the pn interface shifts by 17 nm over 2 V, which results $dW/dV_s = 8.5$ nm/V in close agreement. As a final quantitative comparison, we calculate the average edge-channel separation $W_{\rm avg}$ along the entire interface of setup (II). For this, we approximate the interferometer area of the more complex geometry  and get $W_{\rm avg}=(W_0(L-2L_{sy})+W_s(L_{sx}+2L_{sy}))/(L+L_{sx})$, ($W_0=195$ nm, $W_s=40$ nm: calculated inter-channel separation without and with side gates respectively, and $L_{sx}=500$ nm, $L_{sy}=200$ nm: side gate lengths in the $x$ and $y$-directions respectively). This yields $W_{\rm avg}=$102 nm. The same estimate in the experiments (effective area divided by $L+L_{sx}=1.5 \mu$m) gives $W_{\rm avg}=110$ nm. We find again a very good agreement.

\paragraph*{Conclusion.} The results of this paper show that the edge states structure in a graphene pn junction can be understood quantitatively from the sole knowledge of the device geometry. Besides the implications for our understanding of the underlying physics, this means that conversely, the properties of these interferometers can be engineered. Compared to conventional semiconductors, it opens up new research avenues in electron quantum optics where interaction between propagating edge states can be precisely tuned. This should lead to the demonstration, in future experiments, of more complex quantum operations in graphene such as entanglement \cite{Ionicioiu2001}.

\begin{acknowledgments} I.M.F thanks A. Manesco for interesting discussions.
X.W. acknowledges valuable discussions with  P. Roche, D.C. Glattli and late F. Portier. X.W. acknowledges funding from the FET Open UltrafastNano.
\end{acknowledgments}

\appendix

\bibliography{ref}

\end{document}